\documentclass{emulateapj}
\usepackage{graphicx}
\usepackage{amsmath}
\def\degrees{^\circ}
\def\kms{{\ }{\rm km}\,{\rm s}^{-1}}

\begin{document}
\submitted{Accepted by ApJL on September 7, 2011}

\shortauthors{KAZANTZIDIS ET AL.}
\shorttitle{Formation of dSphs Via Mergers of Disky Dwarfs}

\title{Formation of Dwarf Spheroidal Galaxies Via Mergers of Disky
  Dwarfs}

\author{Stelios Kazantzidis\altaffilmark{1}, 
  Ewa L. {\L}okas\altaffilmark{2},
  Lucio Mayer\altaffilmark{3},
  Alexander Knebe\altaffilmark{4}, and
  Jaros{\l}aw Klimentowski\altaffilmark{2}}

\altaffiltext{1}{Center for Cosmology and Astro-Particle Physics; and
  Department of Physics; and Department of Astronomy, The Ohio State
  University, Columbus, OH 43210, USA; stelios@mps.ohio-state.edu}
\altaffiltext{2}{Nicolaus Copernicus Astronomical Center, 00-716
  Warsaw, Poland} 
\altaffiltext{3}{Institute for Theoretical Physics, University of Z\"urich, 
  CH-8057 Z\"urich, Switzerland} 
\altaffiltext{4}{Grupo de Astrof\'\i sica,
  Departamento de Fisica Teorica, Modulo C-8, Universidad Aut\'onoma
  de Madrid, Cantoblanco E-28049, Spain}

\begin{abstract}
  
We perform collisionless $N$-body simulations to investigate whether
  binary mergers between rotationally-supported dwarfs can lead to the
  formation of dwarf spheroidal galaxies (dSphs). Our simulation
  campaign is based on a hybrid approach combining cosmological
  simulations and controlled numerical experiments. We select merger
  events from a Constrained Local UniversE (CLUES) simulation of the
  Local Group (LG) and record the properties of the interacting
  dwarf-sized halos.  This information is subsequently used to seed
  controlled experiments of binary encounters between dwarf galaxies
  consisting of exponential stellar disks embedded in
  cosmologically-motivated dark matter halos. These simulations are
  designed to reproduce eight cosmological merger events, with initial
  masses of the interacting systems in the range $\sim (5-60) \times
  10^{7} M_{\odot}$, occurring quite early in the history of the LG,
  more than $10$~Gyr ago.  We compute the properties of the merger
  remnants as a distant observer would and demonstrate that at least
  three of the simulated encounters produce systems with kinematic and
  structural properties akin to those of the classic dSphs in the LG.
  Tracing the history of the remnants in the cosmological simulation
  to $z=0$, we find that two dSph-like objects remain isolated at
  distances $\gtrsim 800$~kpc from either the Milky Way or M31. These
  systems constitute plausible counterparts of the remote dSphs Cetus
  and Tucana which reside in the LG outskirts, far from the tidal
  influence of the primary galaxies. We conclude that merging of
  rotationally-supported dwarfs represents a viable mechanism for the
  formation of dSphs in the LG and similar environments.

\end{abstract}


\section{Introduction}
\label{sec:introduction}

The origin of dwarf spheroidal galaxies (dSphs) in the Local Group
(LG) is a long-standing subject of debate. These intriguing systems
are the faintest galaxies known \citep[e.g.,][]{Mateo98} and they are
believed to be highly dark matter (DM) dominated
\citep[e.g.,][]{Mateo98,Lokas09,Walker_etal09}. Moreover, dSphs are
gas poor or completely devoid of gas
\citep[e.g.,][]{Grcevich_Putman09} and they are characterized by
pressure-supported, spheroidal stellar components
\citep[e.g.,][]{Mateo98} and a wide diversity of star formation
histories \citep[e.g.,][]{Grebel00}. While our understanding of dSphs
has grown impressively in the past decade, a definitive model for
their formation still remains elusive.

One class of models attempts to explain the origin of dSphs via
various environmental mechanisms, including tidal and ram pressure
stripping
\citep[e.g.,][]{Einasto_etal74,Faber_Lin83,Mayer_etal01,Kravtsov_etal04,
  Mayer_etal06,Mayer_etal07,Klimentowski_etal09,Kazantzidis_etal11}
and resonant stripping \citep{D'Onghia_etal09}. In this context, the
tidal stirring model \citep{Mayer_etal01} posits the formation of
dSphs via the tidal interactions between rotationally-supported dwarfs
and Milky Way (MW)-sized host galaxies. While this model naturally
explains the tendency of dSphs to be concentrated near the dominant
spirals, it is challenged by the presence of the isolated dSphs Cetus
and Tucana in the LG outskirts.  Such examples illustrate the need for
complementary models concerning the origin of dSphs.

In the current galaxy formation paradigm
\citep[e.g.,][]{White_Rees78}, the quiescent cooling of gas within a
virialized DM halo results in the formation of a
rotationally-supported disk of stars. Thus, any scenario for dSph
formation must incorporate physical processes for transforming
initially rotationally-supported stellar systems to ones dominated by
random motions. Interactions and mergers between galaxies may
constitute such a mechanism.  Indeed, on larger scales, the tidal
heating and violent relaxation associated with mergers of massive,
disk galaxies effectively destroy the stellar disks creating a
kinematically hot, pressure-supported spheroid that resembles an
elliptical galaxy \citep[e.g.,][]{Barnes92}. It has also been
demonstrated that encounters between dwarf halos can lead to their
very strong evolution
\citep[e.g.,][]{Knebe_etal06,Klimentowski_etal10}.
\citet{Klimentowski_etal10} reported in their constrained DM
cosmological simulation of the LG that a few percent of all surviving
subhalos had undergone a substantial encounter with another dwarf halo
in the past.  Most of these events occurred early in the LG history
and before the dwarf halos were accreted and became satellites of
either the MW or M31. Therefore, although the great majority of LG
dwarf galaxies should not have participated in mergers with other
dwarfs in the past, at least some of them may have experienced such
interactions.

Motivated by these findings, we selected eight merger events from the
same LG simulation and re-simulated them at higher resolution,
embedding stellar disks inside the dwarf DM halos.  Our goal is to
determine whether mergers of initially rotationally-supported dwarfs
can produce systems with kinematic and structural properties akin to
those of the classic LG dSphs.


\begin{table*}
\caption{Initial Properties of Merging Dwarfs}
\begin{center}
  \vspace*{-12pt}
\begin{tabular}{lcccccccccccccccc}
\hline
\hline 
\\
\multicolumn{1}{c}{}                   &
\multicolumn{1}{c}{}                   &
\multicolumn{1}{c}{$M_{\rm vir}$}       & 
\multicolumn{1}{c}{$R_{\rm vir}$}       &
\multicolumn{1}{c}{$V_{\rm vir}$}       &
\multicolumn{1}{c}{}                   &
\multicolumn{1}{c}{}                   &
\multicolumn{1}{c}{}                   &                   
\multicolumn{1}{c}{}                   &
\multicolumn{1}{c}{}                   &
\multicolumn{1}{c}{$R_d$}              
\\
\multicolumn{1}{c}{Merger}             &
\multicolumn{1}{c}{$z$}                &
\multicolumn{1}{c}{($10^7 M_{\odot}$)}  & 
\multicolumn{1}{c}{(kpc)}              & 
\multicolumn{1}{c}{(km/s)}             & 
\multicolumn{1}{c}{$c$}                &
\multicolumn{1}{c}{$\lambda$}          &  
\multicolumn{1}{c}{$L_{\rm x}$}         &
\multicolumn{1}{c}{$L_{\rm y}$}         & 
\multicolumn{1}{c}{$L_{\rm z}$}         &
\multicolumn{1}{c}{(pc)}               
\\
\multicolumn{1}{c}{(1)}                &
\multicolumn{1}{c}{(2)}                &
\multicolumn{1}{c}{(3)}                &
\multicolumn{1}{c}{(4)}                &
\multicolumn{1}{c}{(5)}                &
\multicolumn{1}{c}{(6)}                &
\multicolumn{1}{c}{(7)}                &
\multicolumn{1}{c}{(8)}                &
\multicolumn{1}{c}{(9)}                &
\multicolumn{1}{c}{(10)}               &
\multicolumn{1}{c}{(11)}
\\
\\
\hline
\\
M1  &  3.02  &  ~6.51  &  3.15  &  ~9.43   &  4.48  &  0.046 &  $-$0.34  &  ~~0.73    &  $-$0.59   & 119  \\
    &        &  ~5.99  &  3.03  &  ~9.22   &  3.08  &  0.032 &  $-$0.21  &  $-$0.22   &  $-$0.95   & ~97  \\
\\
M2  &  3.82  &  ~7.31  &  2.69  &  10.82   &  4.11  &  0.039 &  $-$0.18  &  $-$0.41   &  $-$0.89   & ~92  \\
    &        &  ~7.17  &  2.67  &  10.75   &  1.92  &  0.033 &  $-$0.70  &  ~~0.68    &  ~~0.21    & 103  \\        
\\
M3  &  6.67  &  ~5.78  &  1.56  &  12.61   &  1.97  &  0.054 &  $-$0.05  &  $-$0.86   &  $-$0.50   & ~90  \\
    &        &  ~7.34  &  1.69  &  13.66   &  1.46  &  0.047 &  $-$0.11  &  $-$0.76   &  $-$0.64   & ~94  \\
\\
M4  &  3.15  &  58.37 &  6.24   &  20.05   &  1.34  &  0.032 &  ~~0.74   &  ~~0.64    &  $-$0.20   & 257  \\
    &        &  19.95 &  4.36   &  14.03   &  1.13  &  0.067 &  ~~0.25   &  ~~0.91    &  $-$0.32   & 340  \\
\\
M5  &  3.82  &  14.76 &  3.41   &  13.65   &  2.92  &  0.034 &  $-$0.68  &  ~~0.13    &  $-$0.72   & 117  \\
    &        &  ~8.21 &  2.79   &  11.26   &  2.61  &  0.065 &  $-$0.37  &  ~~0.81    &  ~~0.47    & 171  \\
\\
M6  &  3.82  &  24.96 &  4.04   &  16.30   &  1.67  &  0.051 &  ~~0.80   &  ~~0.21    &  ~~0.56    & 231  \\ 
    &        &  22.38 &  3.89   &  15.72   &  4.10  &  0.032 &  $-$0.48  &  ~~0.88    &  ~~0.02    & 112  \\
\\
M7  &  2.57  &  ~5.15 &  3.00   &  ~8.59   &  3.09  &  0.045 &  $-$0.54  &  $-$0.79   &  $-$0.31   & 128  \\
    &        &  21.06 &  5.16   &  13.25   &  2.30  &  0.077 &  ~~0.97   &  ~~0.03    &  $-$0.16   & 378  \\
\\
M8  &  3.02  &  41.84 &  5.77   &  17.65   &  6.25  &  0.026 &  ~~0.75   &  $-$0.62   &  $-$0.27   & 116  \\
    &        &  34.18 &  5.39   &  16.52   &  5.88  &  0.031 &  ~~0.99   &  $-$0.07   &  ~~0.09    & 130  \\
\hline
\end{tabular}
\end{center}
\label{table:init_param}
\end{table*}


\section{Methods}
\label{section:methods}

The basis for our experiments is a constrained DM cosmological
simulation of the Local Universe which reproduces the main LG
properties at $z=0$. This simulation
\citep[see][]{Klimentowski_etal10} is part of the CLUES
project\footnote {\texttt{http://www.clues-project.org}}
\citep{Gottloeber_etal10}. We identify dwarf-sized halos participating
in mergers and record their properties by utilizing the halo
catalogues of the cosmological simulation that were generated with the
AMIGA halo finder (AHF) \citep{Knollmann_Knebe09}.

In practice, it is not feasible to simulate all possible encounters
identified in the halo catalogues at sufficient numerical resolution.
For our purposes, it would be sufficient to show that even a single
{\it representative} merger between dwarf galaxies can lead to the
formation of a dSph.  To this end, we selected eight {\it binary}
interactions, which we denote M1-M8, for subsequent re-simulation.
These encounters corresponded to mass ratios of the interacting
systems ranging approximately from $1$:$1$ to $1$:$4$.  While these
merger events were not chosen in any special way, they were associated
with a significant mass transfer between the systems ($\gtrsim 80\%$)
indicating the occurence of an actual merger, as opposed to a weak
encounter or a fly-by.

Table~\ref{table:init_param} contains the properties of the two
interacting systems in each case.  Column 2 lists the redshifts at
which the mergers were identified in the cosmological simulation.  All
recorded encounters took place at very early cosmic times ($z\gtrsim
2.5$), in accordance with the expectation for the epoch of rapid
growth of low-mass halos \citep[e.g.,][]{Wechsler_etal02}. We note
that these mergers occurred before the dwarf halos have been accreted
onto more massive structures. Indeed, subsequent to accretion, the
merging probability of halos diminishes due to the large host velocity
dispersion \citep[e.g.,][]{DeLucia_etal04}.

Columns 3-6 list the halo virial parameters ($M_{\rm vir}$, $R_{\rm
  vir}$, $V_{\rm vir}$) and concentrations $c \equiv R_{\rm vir}/r_s$
(where $r_s$ denotes the ``scale radius'' where the product
$\rho(r)\,r^{2}$ reaches its maximum value). Columns 7-10 list the
dimensionless spin parameters $\lambda$ \citep[e.g.,][]{Peebles69} and
the direction of the angular momentum vector of the dwarf halos
normalized to unity. 

We constructed numerical realizations of dwarfs using the method
described in \citet{Kazantzidis_etal05}. The models consisted of
exponential stellar disks embedded in cosmologically-motivated
\citet{Navarro_etal96} DM halos whose properties match those of the
cosmological halos in Table~\ref{table:init_param}. We parametrized
the mass of the stellar disk as a fraction, $f_d$, of the halo $M_{\rm
  vir}$ and the disk thickness as $z_d/R_d$ ($z_d$ and $R_d$ denote
the disk vertical scale height and radial scale length, respectively).
For simplicity, we fix the values of $z_d/R_d$ and $f_d$ in all
models. In our modeling, $R_d$ is determined by $M_{\rm vir}$,
$\lambda$, $c$, and $f_d$ via the semi-analytic model of
\citet{Mo_etal98} for the formation of disks\footnote{We note that the
  \citet{Mo_etal98} formalism may be inappropriate for dwarf galaxies
  at high-redshift.  However, given the lack of knowledge for the
  structure of dwarf galaxies at early cosmic times, utilizing the
  \citet{Mo_etal98} model constitutes a reasonable alternative to
  assigning arbitrary values to the disk scale lengths of our
  dwarfs.}. The resulting $R_d$ are listed in column 11 of
Table~\ref{table:init_param}.

We take the expectation that dwarf galaxies should be born as thick,
puffy systems \citep[e.g.,][]{Kaufmann_etal07} into account by
conservatively adopting $z_d/R_d = 0.2$. Moreover, we choose $f_d =
0.005$. Such small value is consistent with that inferred for some of
the faintest LG dwarf irregular galaxies (dIrrs) such as Leo A
\citep{Brown_etal07} and SagDIG\citep{Young_Lo97}.  Assuming an
isothermal DM halo, the velocity dispersions of these systems ($\sigma
\sim 10 \kms$) correspond to $V_{\rm vir} \sim 14 \kms$, comparable to
those of our halos.  Tiny stellar components are expected in such
low-mass systems because both gas accretion/retention and star
formation would be strongly suppressed by re-ionization and supernovae
driven outflows \citep[e.g.,][]{Mayer10}.  \citet{Oh_etal11} computed
$f_d$ for a fairly large sample of dwarfs in the THINGS survey,
obtaining values $\sim 0.01$, albeit for galaxies that would have been
less affected by the aforementioned effects having rotational
velocities that are a factor of $2-3$ larger than those considered
here.

We rotated each dwarf model so that the direction of the angular
momentum vector of its DM halo was identical to that of the
corresponding cosmological halo. In our modeling, the angular momentum
of the disk is aligned with that of the host DM halo. This results in
relative orientations of the disk angular momenta that span a wide
range ($\sim 10\degrees-120\degrees$).  Lastly, we realized the merger
simulations using the initial positions and velocities of the
corresponding cosmological halos with respect to their center of mass
(Table~\ref{table:posvel}).

All merger simulations were performed with the $N$-body code PKDGRAV
\citep{Stadel01}. Each $N$-body dwarf model contained $N_h = 3 \times
10^5$ DM particles and $N_d = 5 \times 10^4$ disk particles.  The
gravitational softening was set to $\epsilon_h=30$~pc and
$\epsilon_d=15$~pc, respectively. Force resolution was adequate to
resolve all scales of interest.


\begin{table}
  \caption{Initial Positions and Velocities of Merging Dwarfs}
\begin{center}
  \vspace*{-20pt}
\begin{tabular}{lrrrrrr}
\hline
\hline 
\\
\multicolumn{1}{c}{}                             &
\multicolumn{1}{r}{$x$\,\,\,\,\,}                &
\multicolumn{1}{r}{$y$\,\,\,\,\,}                & 
\multicolumn{1}{r}{$z$\,\,\,\,\,}                &
\multicolumn{1}{r}{$V_{\rm x}$\,\,\,\,\,}         &
\multicolumn{1}{r}{$V_{\rm y}$\,\,\,\,\,}         &
\multicolumn{1}{r}{$V_{\rm z}$\,\,\,\,\,}                   
\\
\multicolumn{1}{c}{Merger}                       &
\multicolumn{1}{r}{(kpc)}                        &
\multicolumn{1}{r}{(kpc)}                        & 
\multicolumn{1}{r}{(kpc)}                        & 
\multicolumn{1}{r}{(km/s)}                       & 
\multicolumn{1}{r}{(km/s)}                       &
\multicolumn{1}{r}{(km/s)}        
\\
\\
\hline
\\
M1  &  $-$16.69  &  $-$17.68   &  $-$4.84    &     4.84    &     1.99    &     0.70   \\
    &  18.15     &     19.22   &     5.26    &  $-$5.26    &  $-$2.16    &  $-$0.76   \\
\\
M2  &  $-$6.67   &     0.29    &  $-$3.13    &     5.52    &     2.36    &     1.73    \\
    &     6.78   &  $-$0.30    &     3.19    &  $-$5.63    &  $-$2.40    &  $-$1.76    \\
\\
M3  &     4.78   &  $-$0.04    &  $-$0.23    &  $-$0.64    &  $-$4.18    &     1.85    \\
    &  $-$3.76   &     0.03    &     0.18    &     0.50    &     3.29    &  $-$1.46    \\
\\
M4  &  $-$1.96   &  $-$3.30    &     0.54    &  $-$3.04    &     0.19    &     0.68    \\
    &     5.74   &     9.65    &  $-$1.58    &     8.89    &  $-$0.56    &  $-$2.00    \\
\\
M5  &  $-$0.02   &     2.06    &     0.18    &   $-$1.10   &   $-$6.38   &     1.74    \\
    &     0.04   &  $-$3.70    &  $-$0.32    &      1.97   &     11.47   &  $-$3.13   \\
\\
M6  &     4.58   &     6.34    &     3.75    &   $-$3.33   &      1.99   &  $-$3.61   \\
    &  $-$5.11   &  $-$7.07    &  $-$4.18    &      3.71   &   $-$2.21   &     4.02   \\
\\
M7  &     1.49   &     2.23    &     6.13    &   $-$0.54   &  $-$20.50   &  $-$5.27   \\
    &  $-$0.36   &  $-$0.57    &  $-$1.50    &      0.13   &      5.01   &     1.29   \\
\\
M8  &  $-$7.76   &  $-$3.02    &  $-$5.87    &      4.54   &   $-$6.91   &  $-$1.21   \\
    &     9.50   &     3.70    &     7.18    &   $-$5.56   &      8.46   &     1.48   \\
\hline
\end{tabular}
\end{center}
\label{table:posvel}
\end{table}



\begin{figure*}[t]
\begin{center}
  \includegraphics[scale=0.3]{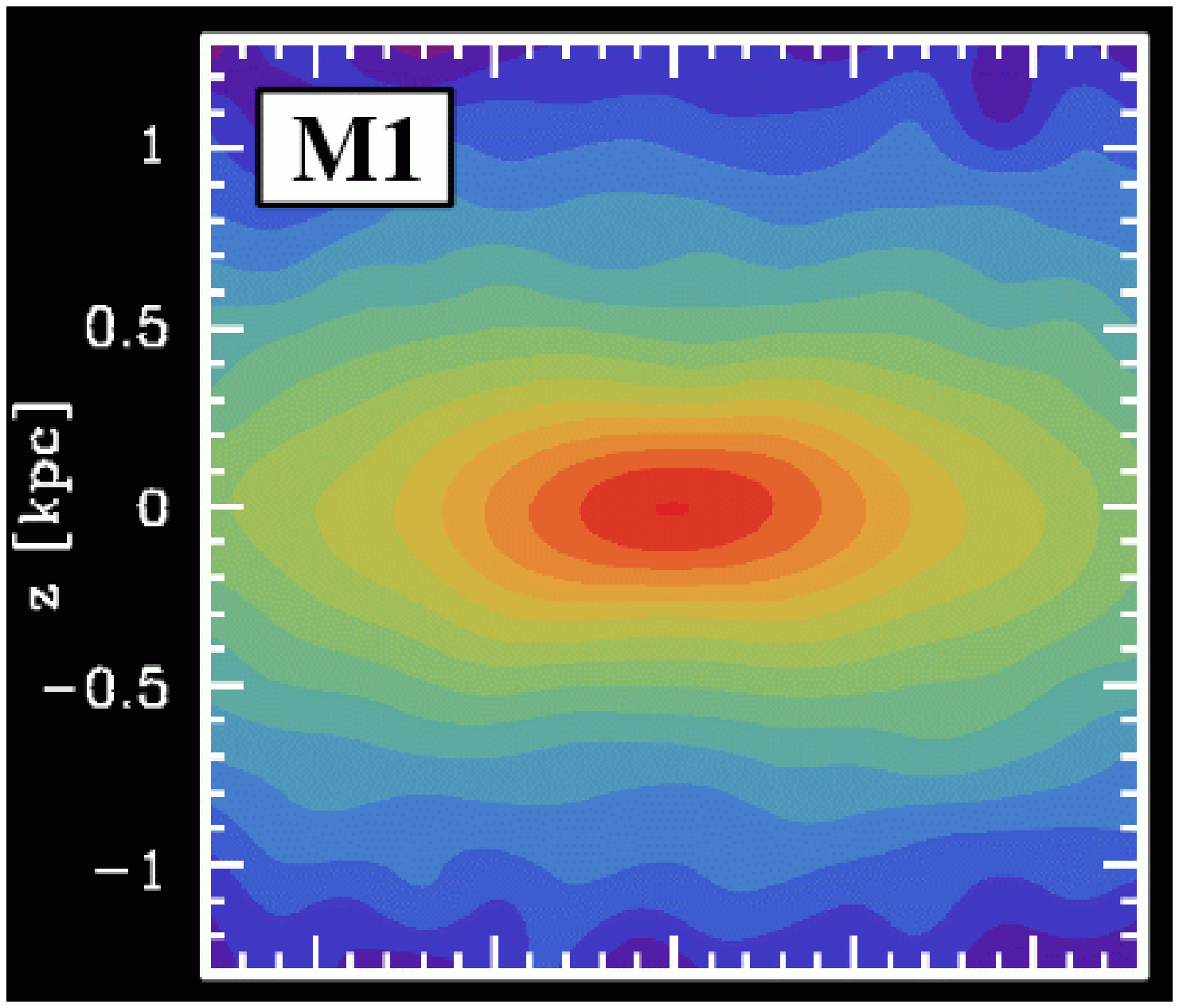}\hspace{-0.1cm}
  \includegraphics[scale=0.3]{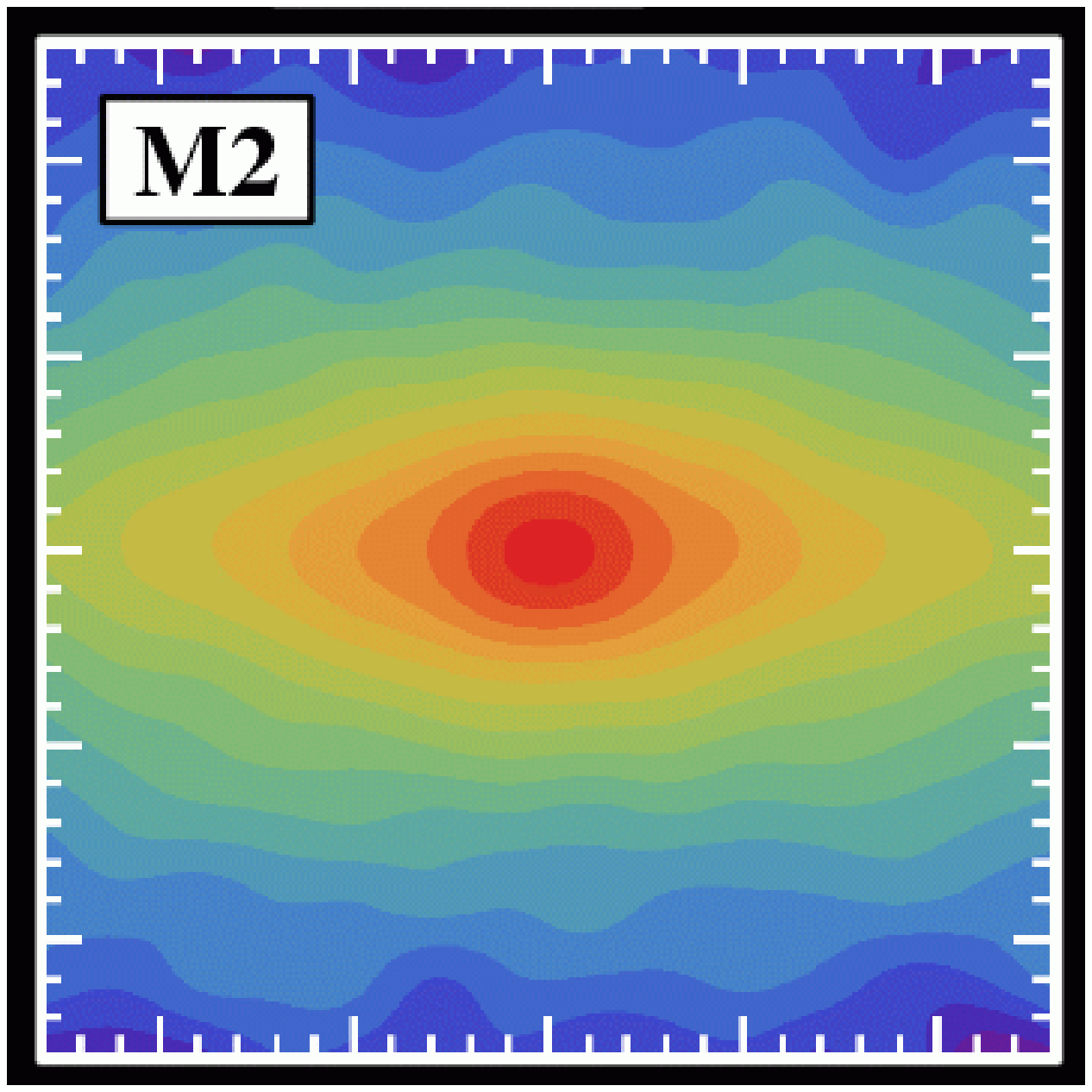}\hspace{-0.1cm}
  \includegraphics[scale=0.3]{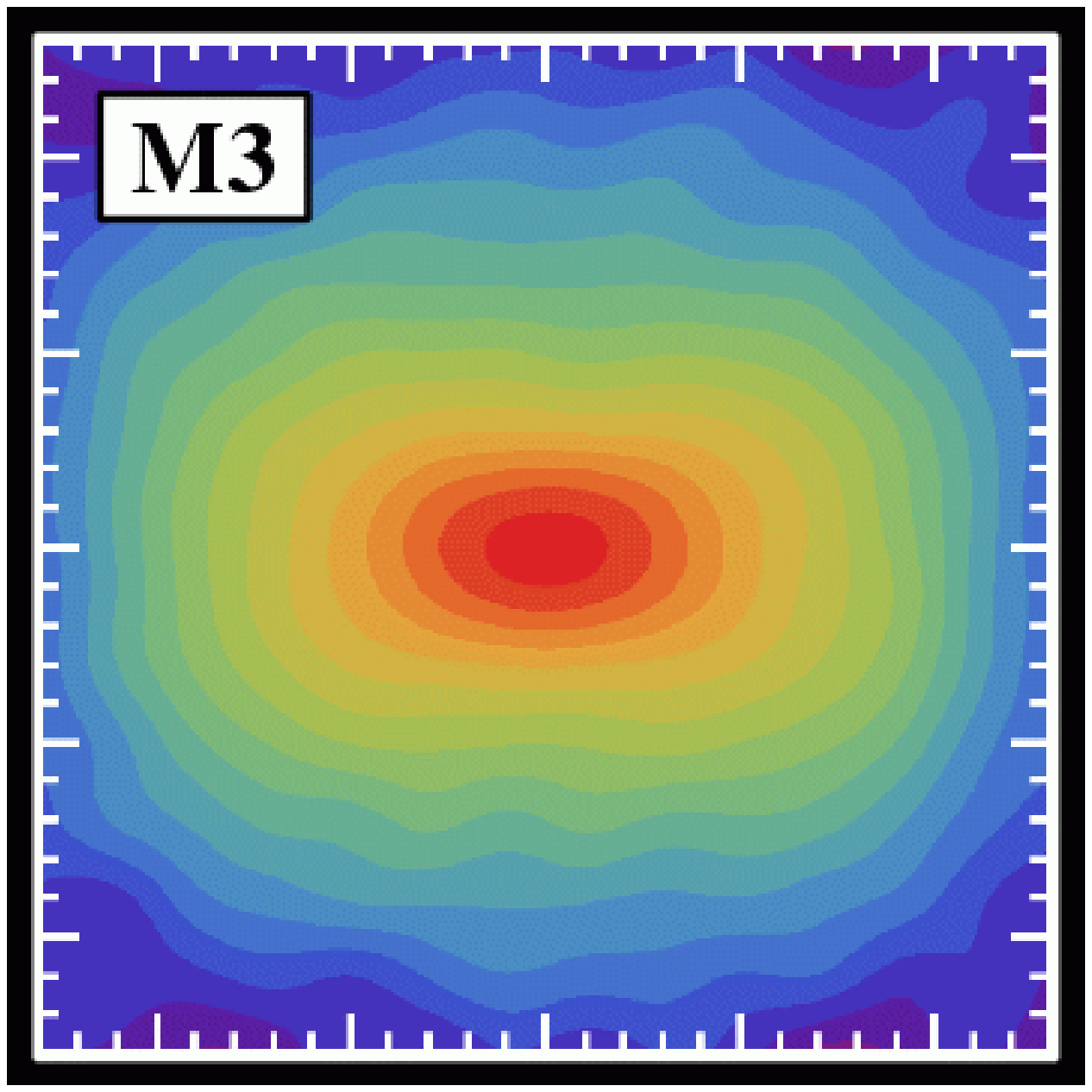}\hspace{-0.1cm}
  \includegraphics[scale=0.3]{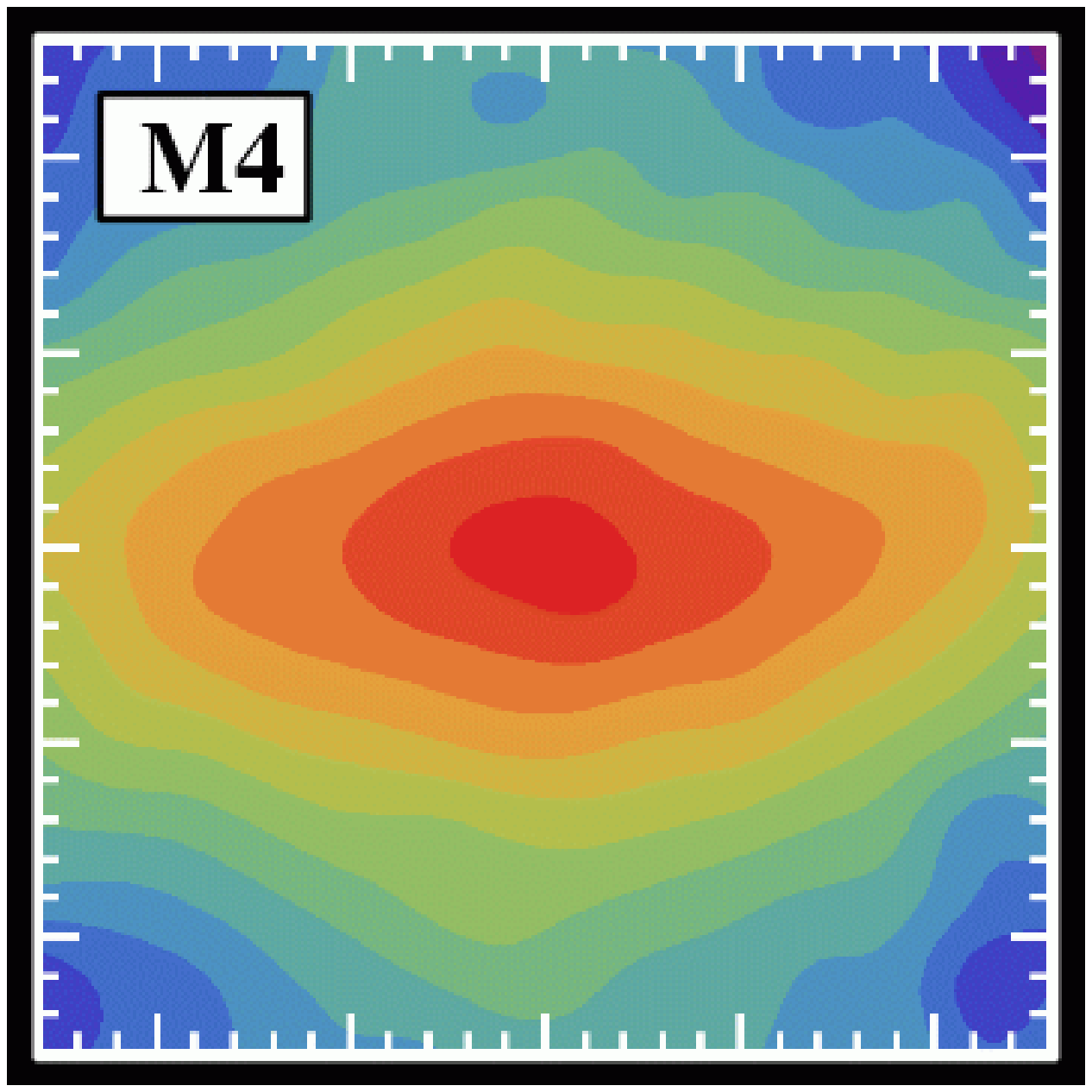}\vspace{-0.1cm}
  \includegraphics[scale=0.3]{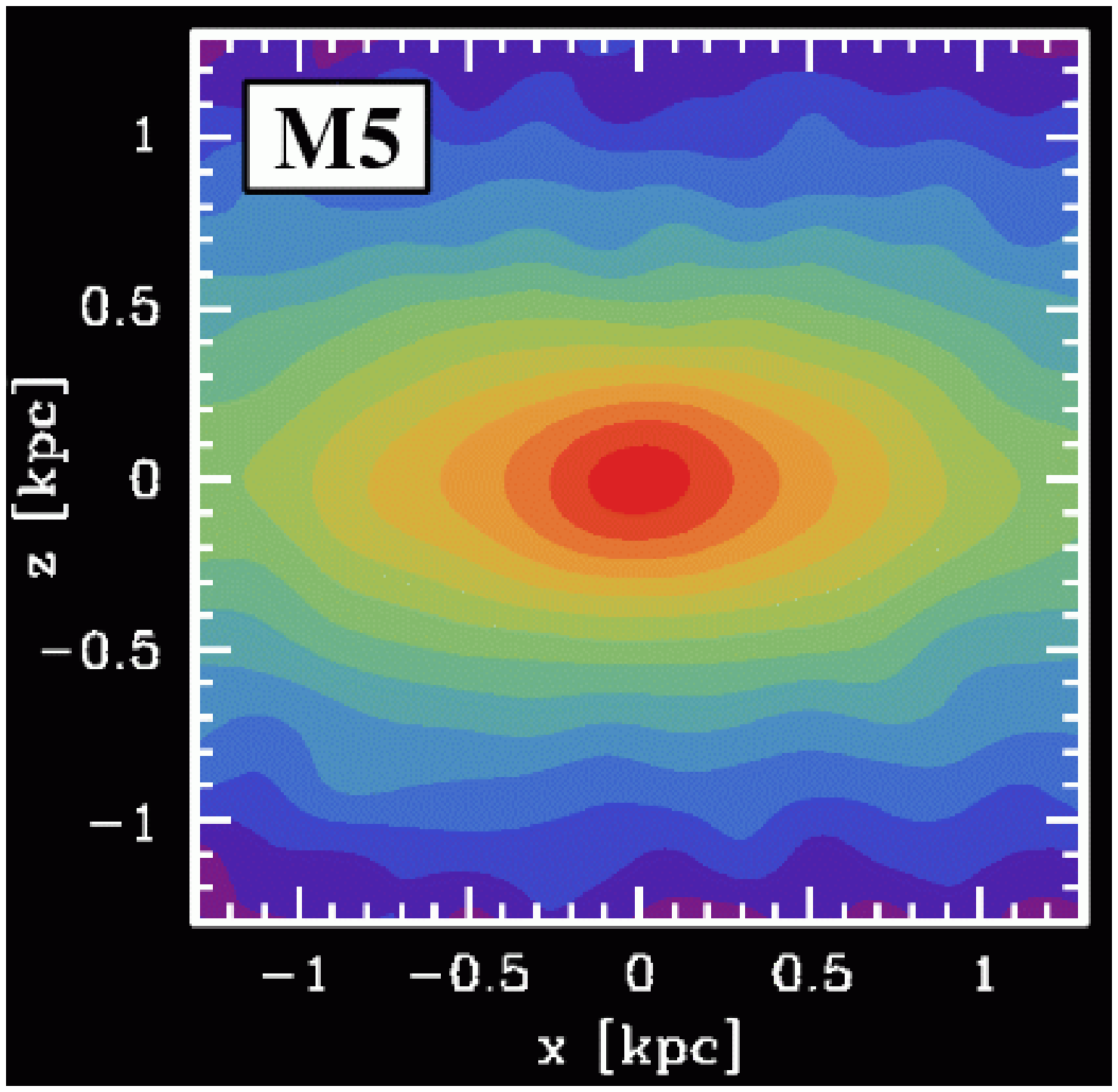}\hspace{-0.1cm}
  \includegraphics[scale=0.3]{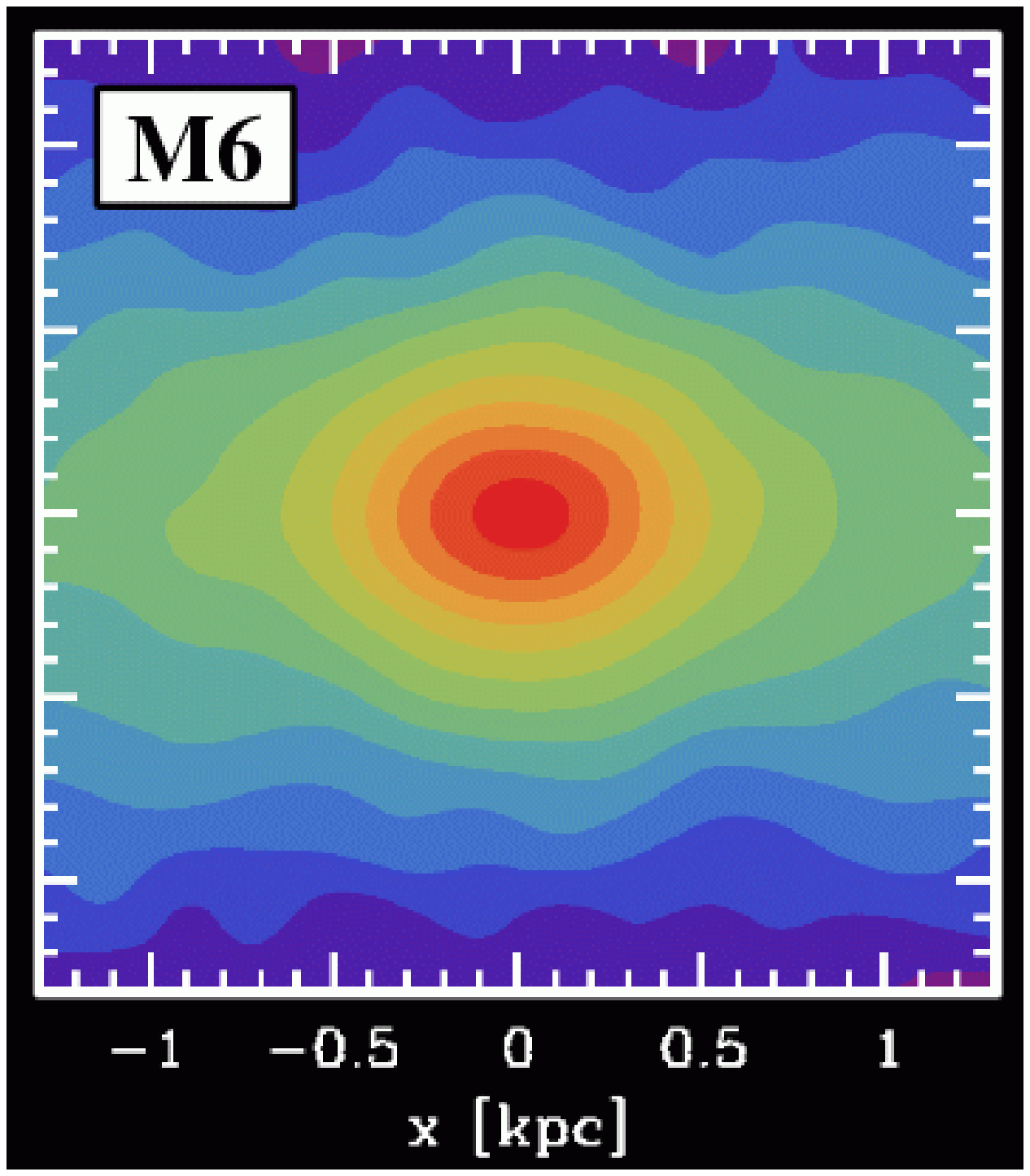}\hspace{-0.1cm}
  \includegraphics[scale=0.3]{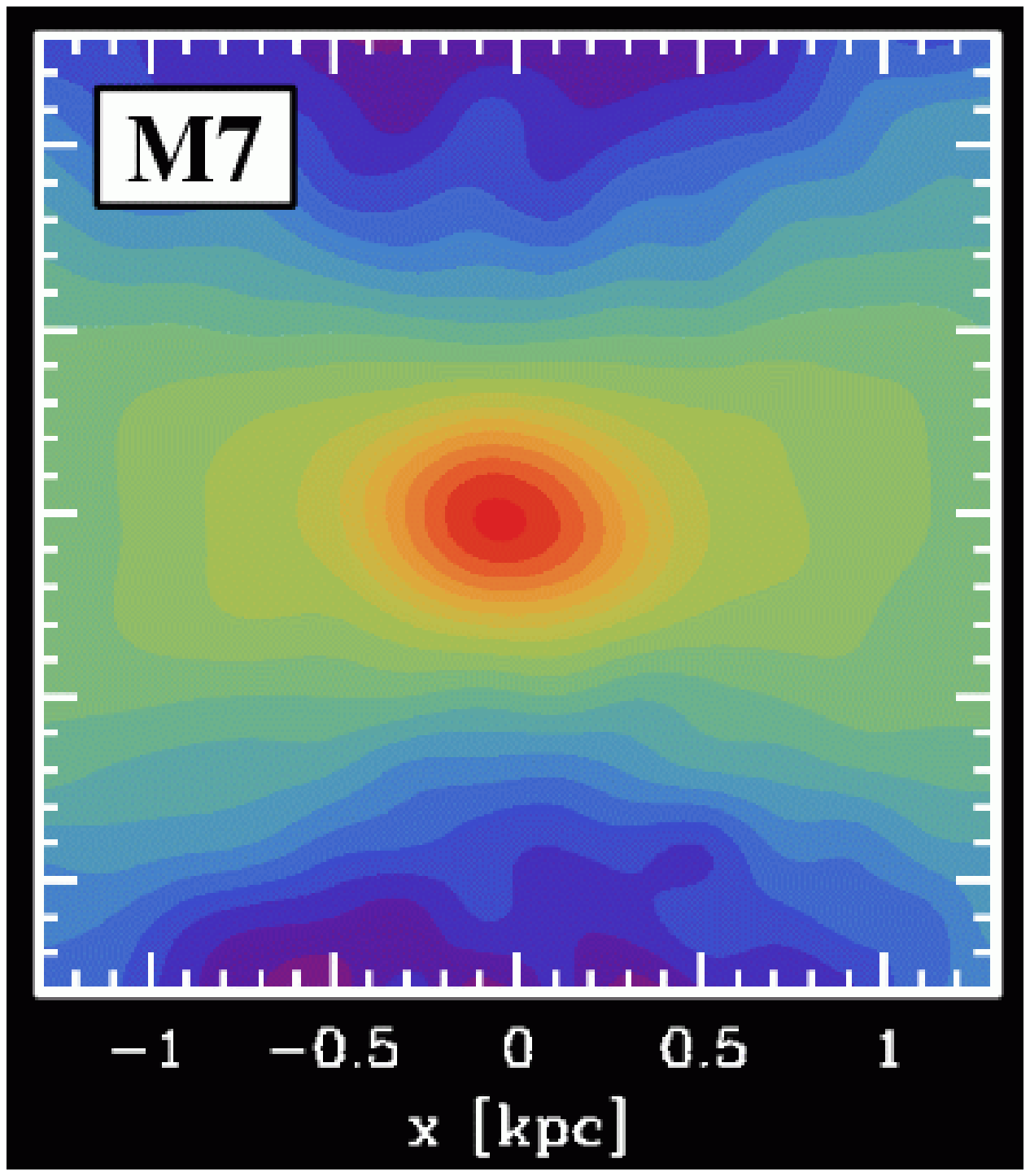}\hspace{-0.1cm}
  \includegraphics[scale=0.3]{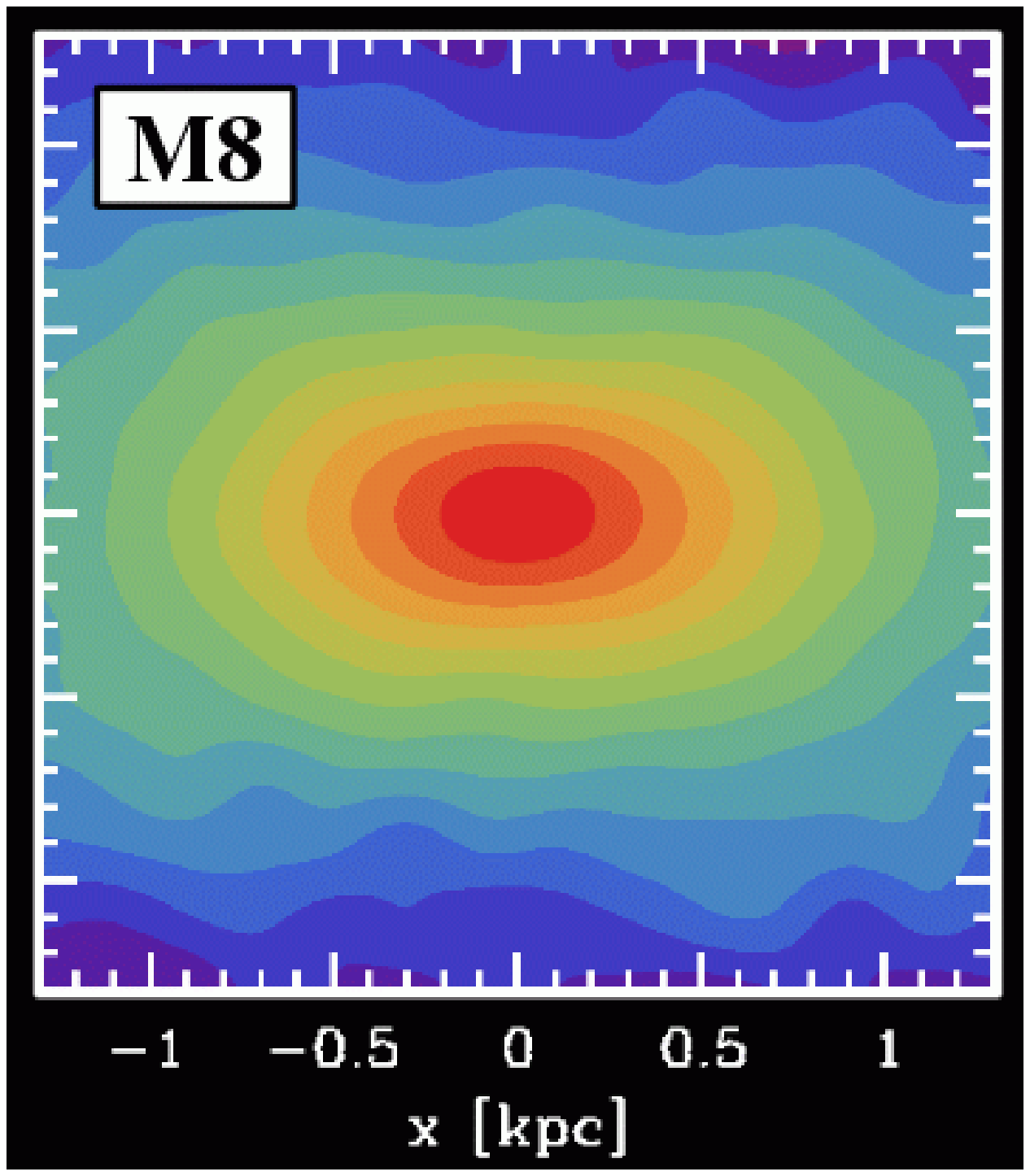}
\end{center}
\caption{Surface number density distribution of stars in the merger 
  remnants. Results are presented for the most non-spherical views
  along the intermediate axis of the stellar distribution, $y$ (where
  $x$ and $z$ denote the major and minor axes, respectively). The
  stars were binned into $0.2~\rm{kpc} \times 0.2~\rm{kpc}$ fields
  perpendicular to the line-of-sight.  The contours correspond to the
  number of stars $N$ within such a bin and are equally spaced by
  $0.2$ in $\log$ N. The innermost contours are in the range $\log
  \rm{N} = 3.2-4.2$ depending on the remnant.
  \label{fig1}}
\end{figure*}



\begin{table*}
  \caption{Properties of Merger Remnants}
\begin{center}
  \vspace*{-12pt}
\begin{tabular}{lcccclcc}
\hline
\hline 
\\
\multicolumn{1}{c}{}                             &
\multicolumn{1}{c}{$r_{1/2}$}                     & 
\multicolumn{1}{c}{}                             &
\multicolumn{1}{c}{}                             &
\multicolumn{1}{c}{}                             &
\multicolumn{1}{c}{$M_V$}                        &
\multicolumn{1}{c}{$\mu_V$ }                     &
\multicolumn{1}{c}{Color}                        
\\
\multicolumn{1}{c}{Merger}                       &
\multicolumn{1}{c}{(kpc)}                        & 
\multicolumn{1}{c}{$V_{\rm rot}/\sigma_{\ast}$}    & 
\multicolumn{1}{c}{$\epsilon \equiv 1-b/a$}      &
\multicolumn{1}{c}{Classification}               &
\multicolumn{1}{c}{(mag)}                        &
\multicolumn{1}{c}{(mag arcsec$^{-2}$)}           & 
\multicolumn{1}{c}{in Figure~\ref{fig2}}         
\\
\multicolumn{1}{c}{(1)}                          &
\multicolumn{1}{c}{(2)}                          &
\multicolumn{1}{c}{(3)}                          &
\multicolumn{1}{c}{(4)}                          &
\multicolumn{1}{c}{(5)}                          &
\multicolumn{1}{c}{(6)}                          &
\multicolumn{1}{c}{(7)}                          & 
\multicolumn{1}{c}{(8)}       
\\
\\
\hline
\\
M1  &  0.22-0.40  &  0.10-0.20  &  0.11-0.63  &  dSph?     &  $-$(8.3-9.2)    &  25.1-27.1  &  green   \\
M2  &  0.22-0.38  &  0.09-0.69  &  0.04-0.59  &  dSph?     &  $-$(8.5-9.4)    &  24.9-26.6  &  red     \\ 
M3  &  0.24-0.32  &  0.12-0.70  &  0.14-0.45  &  dSph      &  $-$(8.4-9.3)    &  25.0-26.7  &  blue    \\ 
M4  &  1.24-1.50  &  0.19-1.88  &  0.12-0.50  &  non-dSph  &  $-$(10.3-11.2)  &  26.4-27.6  &  black   \\ 
M5  &  0.30-0.52  &  0.20-0.60  &  0.03-0.59  &  dSph?     &  $-$(9.0-9.9)    &  25.1-26.8  &  orange  \\ 
M6  &  0.42-0.60  &  0.22-0.75  &  0.06-0.50  &  dSph      &  $-$(9.7-10.7)   &  24.7-26.1  &  magenta \\  
M7  &  0.28-0.38  &  0.33-0.92  &  0.08-0.54  &  dSph?     &  $-$(9.1-10.0)   &  24.4-25.9  &  cyan    \\ 
M8  &  0.29-0.41  &  0.09-0.50  &  0.07-0.47  &  dSph      &  $-$(10.3-11.2)  &  23.7-25.3  &  brown   \\ 
\hline
\end{tabular}
\end{center}
\label{table:dSphs}
\end{table*}


\section{Results}
\label{sec:results}

All merger products were allowed to reach equilibrium before any
analysis was performed. We first determined the principal axes of the
stellar components using the moments of the inertia tensor.
Subsequently, we produced 2D maps of the stellar surface distribution
and kinematics along the three principal axes and ``observed'' the
merger remnants as a distant observer would.

Given that observed dSphs are supported by random motions
\citep[e.g.,][]{Mateo98}, only remnants satisfying $V_{\rm
  rot}/\sigma_{\ast} \lesssim 1$ (where $V_{\rm rot}$ and
$\sigma_{\ast}$ denote the stellar rotational velocity and the
one-dimensional, central velocity dispersion, respectively) may be
regarded as dSphs.  Using the kinematic maps of the stellar mean
radial velocity and velocity dispersion, we computed $V_{\rm
  rot}/\sigma_{\ast}$ for all remnants.  For each line-of-sight, the
value of $V_{\rm rot}$ corresponded to the maximum velocity found
anywhere on the map. For the velocity dispersion $\sigma_{\ast}$ we
adopted the central value.

Most classic dSphs have projected ellipticities, $\epsilon \equiv
1-b/a$ (where $b$ and $a$ denote the projected minor and major axis of
the stellar distribution, respectively) in the range $0.1 \lesssim
\epsilon \lesssim 0.5$ \citep[e.g.,][]{Mateo98,McGaugh_Wolf10}.
Therefore, we classify as dSphs only those remnants that satisfy
$\epsilon \lesssim 0.5$. For each line-of-sight, we measured
$\epsilon$ at a distance of $2\,r_{1/2}$, where $r_{1/2}$ is the
half-light radius.  Such distances are consistent with those adopted
in observational studies \citep[e.g.,][]{McConnachie_Irwin06}.
$r_{1/2}$ were determined by calculating the radius containing half
the total number of stars in the surface density distribution of each
remnant.  The range of $r_{1/2}$ in Table~\ref{table:dSphs}
corresponds to the minimum and maximum values determined for the lines
of sight along the three principal axes and reflects the fact that the
remnants are not spherical.  Figure~\ref{fig1} shows the surface
density maps of all eight merger remnants.

Columns 3-5 of Table~\ref{table:dSphs} list the range of values of
$V_{\rm rot}/\sigma_{\ast}$ and $\epsilon$ for each remnant together
with the outcome of the classification scheme based on the two
criteria above. Any random line-of-sight would correspond to values of
$V_{\rm rot}/\sigma_{\ast}$ and $\epsilon$ within the ranges quoted in
Table~\ref{table:dSphs}.

Out of eight initial merger remnants, three (M3, M6, and M8) would be
classified as bona fide dSphs. If we slightly relax the shape
criterion above by considering the most elongated classic dSph Ursa
Minor with $\epsilon \approx 0.6$ \citep{Mateo98}, then seven of our
remnants would qualify as dSphs (we indicate this by using the
notation "dSph?" in column 5 of Table~\ref{table:dSphs} to
characterize remnants M1, M2, M5, and M7). As in mergers between
massive disk galaxies \citep[e.g.,][]{Cox_etal06}, the values of
$V_{\rm rot}/\sigma_{\ast}$ are affected by both the mass ratios of
the merging systems and the degree of initial alignment of the disks'
angular momenta. Indeed, despite the near alignment of the disk spins
in M8, a dSph does form. This is because the disks are effectively
destroyed in this nearly equal-mass encounter.  On the other hand,
mergers M4 and M7 which correspond to the smallest mass ratios produce
remnants that exhibit the highest values of $V_{\rm
  rot}/\sigma_{\ast}$. In these cases, the remnants maintain part of
the original rotation of the more massive disks which were not
completely destroyed by the encounters. Moreover, although M7
corresponds to a smaller mass ratio compared to M4, its remnant
exhibits a lower value of $V_{\rm rot}/\sigma_{\ast}$ and is
classified as a dSph. This is because the intrinsic spins of the disks
in this case are oriented almost in the opposite direction, whereas
the disk spins are nearly aligned in M4.

Columns 6 and 7 of Table~\ref{table:dSphs} list the absolute magnitude
in the V-band, $M_V$, and the central surface brightness, $\mu_V$. For
simplicity, the magnitudes were determined by assuming that our
imaginary observer will be able to detect all stars. The total stellar
masses were translated into luminosities assuming a stellar
mass-to-light ratio of $M_{\ast}/L_V = (2.5 \pm 1)
M_{\odot}/L_{\odot}$. The mean value is the {\it present-time}
prediction for a single, low-metallicity stellar population in the
standard model described by \citet{Bruzual_Charlot03}. The lower limit
roughly corresponds to the mean $M_{\ast}/L_V$ for pressure-supported
LG dwarfs estimated from resolved stellar population studies
\citep[e.g.,][]{Mateo98}. The $1 M_{\odot}/L_{\odot}$ variation
addresses some of the uncertainty regarding our assumptions for the
mean $M_{\ast}/L_V$ above and the lack of relevant physical processes
in the simulations (e.g., star formation) and illustrates how the
predicted magnitudes may vary due to unknown $M_{\ast}/L_V$.

Lastly, $\mu_V$ were computed by considering stars within
$0.2\,r_{1/2}$ from the center of each remnant.  Such distances
roughly correspond to the innermost data points of surface brightness
distributions of observed dwarfs \citep[e.g.,][]{McConnachie_Irwin06}.
The ranges of $\mu_V$ in column 7 of Table~\ref{table:dSphs} include
both the variation of $r_{1/2}$ and the uncertainty in $M_{\ast}/L_V$.


\begin{figure}
  \begin{center}
    \includegraphics[scale=1.2]{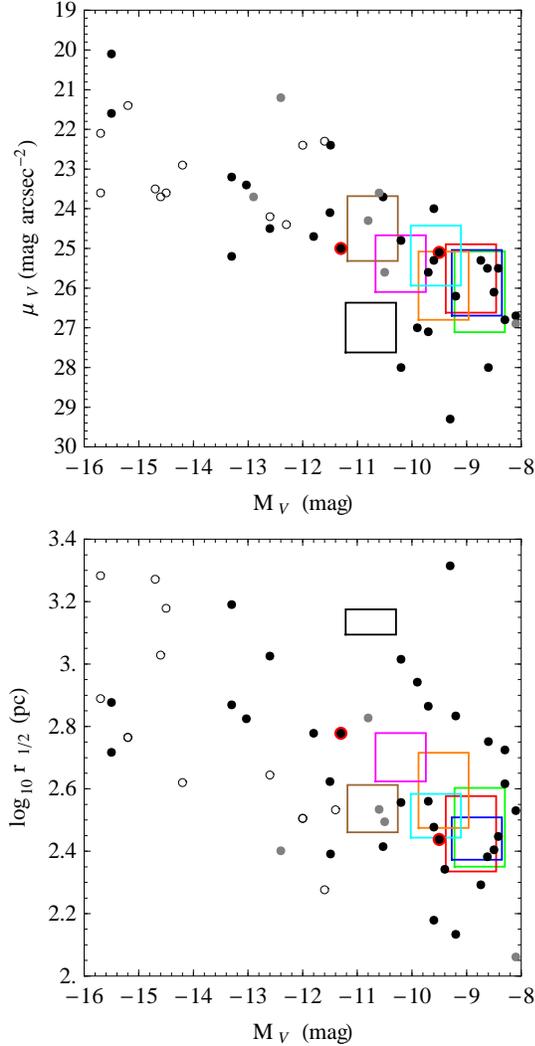}
\end{center}
\caption{Absolute magnitude, $M_V$, versus central surface brightness,
  $\mu_V$ (upper panel) and $M_V$ versus projected half-light radius,
  $r_{1/2}$ (bottom panel) for dwarf galaxies. The merger remnants correspond to color
  rectangles whose sizes indicate the ranges of values in
  Table~\ref{table:dSphs}.  Symbols show results for observed LG
  dwarfs: open symbols correspond to dIrrs,
  filled black symbols show results for both dSph and dwarf
  spheroidal/dwarf elliptical (dSph/dE) systems, and filled gray
  symbols correspond to ``transition-type'' (dIrr/dSph) dwarfs. 
  Isolated dSphs Cetus and Tucana are additionally marked with red circles.
  All data for observed dwarfs are taken from Table~2 of
  \citet{Lokas_etal11}. The remnants that would be classified 
  as dSphs and the LG classic dSphs occupy essentially the same regions 
  on both planes.
  \label{fig2}}
\end{figure}


\section{Discussion}
\label{sec:discussion}

Galaxies can be characterized by correlations between their
fundamental parameters. This also applies to LG dwarf galaxies which
occupy characteristic regions in the $M_V-\mu_V$ and $M_V-r_{1/2}$
planes \citep[e.g.,][]{Tolstoy_etal09}.  Figure~\ref{fig2} illustrates
how our merger remnants (colored rectangles) compare with the
population of LG dwarf galaxies (symbols) in these two planes.  Except
M4 which is not classified as a dSph, our merger products and the LG
classic dSphs (filled black symbols) occupy essentially the same
regions on the $M_V-\mu_V$ and $M_V-r_{1/2}$ planes. This further
suggests that merging between rotationally-supported dwarfs
constitutes a viable mechanism for the formation of dSphs.

Using the subhalo catalogues we have traced the history of the merger
remnants in the cosmological simulation to $z=0$. From the seven
systems that could be classified as dSphs, four (M2, M3, M6, and M7)
survived as individual entities to the present-time. Two of the
remaining merger products (M1 and M8) were accreted by a factor of
$\sim 100$ more massive halos, while M5 participated in a $1$:$5$
interaction with a more massive object.  Interestingly, all four
merger remnants that have survived as individual entities to $z=0$
appear to be remote, located in the periphery of the simulated LG at
distances $\gtrsim 800$~kpc from either the MW or M31.  These systems
constitute plausible counterparts of the oddly isolated dSphs Cetus
and Tucana which reside in the LG outskirts, the former at a distance
of $\sim 800$~kpc from the MW \citep[e.g.,][]{McConnachie_etal05}
while the latter at $\sim 1300$~kpc from M31
\citep[e.g.,][]{Saviane_etal96}. The observational parameters of
Tucana ($M_V = -9.5$~mag, $r_{1/2} = 274$~pc, $\mu_V = 25.1$ mag
arcsec$^{-2}$, $V_{\rm rot}/\sigma_{\ast} \sim 1$, and $\epsilon =
0.48$; Table~2 of \citealt{Lokas_etal11}) are in reasonable agreement
with those of dSphs M3 and M8 in Table~\ref{table:dSphs}\footnote{Due
  to the very large distances involved the detection of rotation at
  such high levels for Tucana should be regarded as tentative and
  needs to be confirmed by future observations.}. On the other hand,
the Cetus dwarf ($M_V = -11.3$~mag, $r_{1/2} = 600$~pc, $\mu_V = 25.0$
mag arcsec$^{-2}$, $V_{\rm rot}/\sigma_{\ast} = 0.45$, and $\epsilon =
0.33$; Table~2 of \citealt{Lokas_etal11}) is brighter and more
extended and therefore more akin to our dSph M6. Overall, the
agreement is particularly noteworthy because our simulation program
did not aim to reproduce the properties of these objects.

The evolutionary scenario proposed here may also be relevant for the
relatively isolated transitional (dIrr/dSph) dwarfs Phoenix, LGS3, and
Pegasus \citep{Mateo98} that usually differ from dSphs only in the
presence of small amounts of gas, and even for the remote peculiar
dwarf galaxy VV124 \citep{Kopylov_etal08}.

Alternative models exist for explaining the puzzling presence of
Tucana and Cetus in the outskirts of the LG.  For example, as a result
of three-body interactions, satellites can acquire extremely energetic
orbits with apocenters beyond the $R_{\rm vir}$ of the primary and be
ejected to large distances \citep{Sales_etal07,Ludlow_etal09}. In this
model, ejected subhalos are typically the least massive members of a
group of satellites that is tidally disrupted by the host. Tidal
interactions with the more massive companions within the groups may
already transform the eventually ejected dwarfs into dSphs, via either
tidal stirring, perhaps enhanced by resonances
\citep{D'Onghia_etal09}, or mergers. Ongoing investigations of the
stellar components of Tucana and Cetus
\citep[e.g.,][]{Bernard_etal09,Monelli_etal10a}, including their star
formation histories \citep[e.g.,][]{Monelli_etal10b}, may soon provide
constraints on the various competing models for their origin.  

Lastly, our simulations neglect gas dissipation and star formation.
Dwarf galaxies are known to be gas rich at least at low redshift
\citep[e.g.,][]{Geha_etal06}. Although such dissipative processes
should be important in encounters of massive galaxies with cold gas
disks \citep[e.g.,][]{Kazantzidis_etal05}, they should have a rather
marginal effect in our case. Indeed, the dwarf halos considered here
(Table~\ref{table:init_param}) have both low $V_{\rm vir} < 20\kms$
and concentrations ($ 1 \lesssim c \lesssim 6.5$), implying maximum
circular velocities $V_{\rm max}$ only slightly above $V_{\rm vir}$.
\citet{Kaufmann_etal07} have shown that halos with $V_{\rm max} < 35
\kms$ would not host thin gaseous disks but rather spheroidal gas
distributions.

Furthermore, neglecting self-shielding, Mayer et al. (2006) have shown
that the gas in dwarf halos would still be heated to $T > 10^4$~K by
the cosmic ionizing background at $z > 2$, when all of our mergers
occur.  Given that $V_{\rm vir} \sim 20 \kms$ corresponds to $T_{\rm
  vir} \sim 10^4$~K, the gas distribution would form an extended
pressure-supported envelope.Depending on halo mass, this envelope
would be barely confined by the halo potential, or even be completely
evaporated from the halo \citep{Barkana_Loeb99}.  Regardless, the gas
component would not remain confined into the central disk, except
perhaps near the very center where self-shielding might be sufficient
to preserve a tiny fraction of cold dense gas \citep{Susa_Umemura04}.
Therefore, either the mergers would occur essentially in a
dissipationless regime, as we have assumed here, or they would contain
a tenuous hot atmosphere which could be shock heated further as a
result of the merger \citep[e.g.,][]{Kazantzidis_etal05} and become
unbound thereafter in the absence of efficient cooling. In either
case, no gas inflows and no appreciable triggered star formation would
occur, contrary to what happens in mergers of massive galaxies
\citep[e.g,.][]{Barnes_Hernquist96}. In summary, the effects of gas
and star formation on the structure of the merger remnants are
expected to be weak. Therefore, although they will have to be
confirmed with hydrodynamical simulations, the qualitative arguments
outlined above suggest that our collisionless experiments have likely
captured the essence of the morphological evolution in mergers of
disky dwarf galaxies.

\acknowledgments

The authors would like to thank the referee, Matteo Monelli, for
constructive comments on the manuscript and the CLUES members S.
Gottl\"{o}ber, Y.  Hoffman, and G. Yepes for providing the LG
simulation. We also acknowledge stimulating discussions with J\"urg
Diemand, Alan McConnachie, Chris Orban, and David Weinberg. S.K. is
supported by the Center for Cosmology and Astro-Particle Physics at
The Ohio State University.  This research was partially supported by
the Polish National Science Centre under grant N N203 580940. A.K. is
supported by the Ministerio de Ciencia e Innovacion (MICINN) in Spain
through the Ramon y Cajal program and further acknowledges support
from grants AYA 2009-13875-C03-02, AYA2009-12792-C03-03, and CAM
S2009/ESP-1496. This research was also supported by the Ohio
Supercomputer Center (http://www.osc.edu).

\bibliography{ms} 

\end{document}